\documentclass[12pt]{article}

\usepackage[all]{xy}

\usepackage[american]{}
\usepackage[ansinew]{inputenc}
\usepackage{amsfonts}
\usepackage{amsthm,amssymb,amscd,amsbsy}
\usepackage{amsmath}
\usepackage{epsfig}
\usepackage{epstopdf}
\usepackage[usenames,dvipsnames]{color} 
\usepackage{graphicx}
\usepackage{graphics}
\usepackage{hyperref}
\usepackage{url}

\newcommand{\EE}{\mathbb{E} }

\newcommand{\PP}{\mathbb{P} }

\newcommand{\bi}{\begin{itemize}}
\newcommand{\ei}{\end{itemize}}
\newcommand{\beq}{\begin{equation}}
\newcommand{\eeq}{\end{equation}}
\newcommand{\beas}{\begin{eqnarray*}}
\newcommand{\eeas}{\end{eqnarray*}}
\newcommand{\bea}{\begin{eqnarray}}
\newcommand{\eea}{\end{eqnarray}}
\newcommand{\bv}{\begin{verbatim}}
\newcommand{\ev}{\end{verbatim}}

\newtheorem{remark}{Remark}[section]

\numberwithin{equation}{section}

\hypersetup{
backref=true,
pagebackref=true,
hyperindex=true,
linkcolor=blue,
bookmarks=true,
pdftoolbar=true,
pdfmenubar=true,
bookmarksopen=true,
breaklinks=true,
urlcolor= blue,
citecolor=green,
filecolor=magenta,
pagebackref=true,
pdftitle={Market impact and optimal market making in presence of order flow correlations},
pdfauthor={Jonathan Donier},
}

\title{Market Impact with Autocorrelated Order Flow under Perfect Competition}
\date{December 17, 2012}

\author{Jonathan Donier$^{\dag}$$^{\ast}$}

\begin{document}

\maketitle

\begin{center}
$^{\dag}$ \textit{Ecole Polytechnique, Paris.} {\tt  jonathan.donier@polytechnique.org}\\
$^{\ast}$ \textit{written under the supervision of Jim Gatheral, Department of Mathematics, Baruch College, CUNY.}
\end{center}

\vspace*{5mm}

\begin{center}
{\large \textbf{PRELIMINARY DRAFT - COMMENTS APPRECIATED}}
\end{center}

\vspace*{15mm}

\begingroup
\hypersetup{ pdfborder = {0 0 0}}
\tableofcontents
\endgroup

\vspace*{15mm}

\begin{abstract}

Our goal in this paper is to study the market impact in a market in which order flow is autocorrelated. We build a model which explains qualitatively and quantitatively the empirical facts observed so far concerning market impact. We define different notions of market impact, and show how they lead to the different price paths observed in the literature. For each one, under the assumption of perfect competition and information, we derive and explain the relationships between the correlations in the order flow, the shape of the market impact function while a meta-order is being executed, and the expected price after the completion. We also derive an expression for the decay of market impact after a trade, and show how it can result in a better liquidation strategy for an informed trader. We show how, in spite of auto-correlation in order-flow, prices can be martingales, and how price manipulation is ruled out even though the bare impact function is concave. We finally assess the cost of market impact and try to make a step towards optimal strategies.

\end{abstract}

\section{Introduction}

Market impact has been studied in more and more detail lately, and its understanding is important for different reasons. If one considers the market from the outside, the market impact represents the quantity of information conveyed by the trades - the information being summarized in the price. Its understanding is therefore a step towards the understanding of the price process, of its stability - or instability - and its causes. If on the contrary one considers the market from the inside, the market impact represents the impact of our trade on the price, ie the fundamental cost of executing an order. This cost is not fixed a priori, and should depend on the execution strategy, ie the way one submits the order while other trades occur. We have already described here two types of market impacts: in order to be clear throughout the paper, let us give some definitions first.

The \textit{Bare Market Impact} is the impact of the orders sent by the whole market. It is the most fundamental and universal quantity, since the associated phenomenon can be modeled and its understanding will allow us to understand all the other types of market impact. Let us explain. Market impact is equivalently the way the market reacts to trades: but since the market does not know who sends the orders, it can only respond to the whole order flow - this, and only this, can be modeled and give us the bare market impact. When, like here, one looks at the aggregation of all the trades, there is no ambiguity regarding the definition, since the price path is equal to the impact. This definition may be useful if one wants to study the market as a whole, the price process, or simply find the appropriate response to trades, etc. But an actor inside the market is essentially interested in his own impact, which leads to our next definitions.

The \textit{Renormalized Market Impact} is the impact of a trade submitted by one actor, ie the shift in the price which comes from his trade. This must not be mistaken with the \textit{Observed Market Impact}, which is the price at which the actor trades - imagine for instance that an actor buy while there is a downwards trend : the observed market impact would then be negative but he would still have a positive impact on the price.

Let us make a step further in these definitions. The observed phenomenon is the following: when a buy meta order is sent, its immediate effect is to move the price upwards, and it is observed that after completion the prices decreases to a price that we will call $p_{\infty}$. The interpretation is simple. Obviously, the immediate effect of a buy order is to push the price up, both because it is a positive signal saying that the price is too low, and because it eats the orders on only one side of the order-book. The decrease observed afterwards will be described in detail later, and can be seen as a refill on the depleted side of the order book. This leads to two new definitions.

The \textit{Transient Bare Market Impact} is the price path while information is conveyed to the market - or during what we will call a trend.

The \textit{Permanent Bare Market Impact} is the impact which remains after the decay. Ie, if while the market is in equilibrium - as many buy and sell orders of any kind - an actor sends buy orders, the price while he trades will be equal to the transient bare market impact and after completion it will decrease to the permanent bare market impact. This virtual actor represents in reality the aggregation of actors who participate to the trend. In the same way, one can define these quantities for individual trades (ie the renormalized impact).

Whereas the bare market impact is universal, the observed market impact depends on the strategy. Therefore, one must be careful when one analyzes the observed market impact. For example, we will see that according to the quality of the strategy, there may be no decrease after completion, or the observed permanent impact can be zero or even negative. In the first case, corresponding to the trend initiator, the execution price is lower than the average, whereas in the second case, corresponding to the trend follower, the execution price is higher than the average : this raises the issue of information, and of the relevance of betting with no information - or late information. This will be studied quantitatively later.

\begin{remark}
In the whole paper, we study the impact of the information contained only in the trades - that is to say, private information. The case of public information which is received simultaneously by all the actors is different, and should result in a jump with little or few transactions.
\end{remark}


\section{Representation of the market}

We consider a market in which there are correlations in the order flow (it is essentially found to be power-law, as shown in \cite{eisler}). We want to understand how the market reacts to these trades, which represent some information (public or private), assuming perfect competition for the actors who react to the trades (giving rise to a zero-profit condition, as for the market maker in \cite{glosten} for example). Understanding this reaction will by definition allow one to find the expression of the market impact of trades, for each definition of the impact.\\

Contrary to what is sometimes done in the literature, we do not need to introduce different agents with distinct behaviors to find an equilibrium (which would lead to conceptual problems, as emphasized in \cite{farmer5}: one could hardly imagine that there could be any competition between perfectly intelligent agents and agents with no intelligence at all in financial markets). Our way to decompose the market does not take into account the individual participants, but is purely statistical. We decompose the order-flow in two components: the \textit{signal} (which we will call informed trades) and the \textit{noise} (which we will call uninformed trades). Note that these two components do not represent actual agents: each trader could contribute to both. A personification is made for convenience; this allows us to define the following representation of the market:

\begin{itemize}
\item The informed trader generates autocorrelation in the order sign process (order flow for short) according to the order-splitting model of \cite{farmer2}. This virtual actor represents the current trend in the whole market, and thus an aggregation of trades from possibly many different actors. His existence comes from the fact that long-ranged information is incorporated in the market (especially information that is not available to all the actors, so that for the other actors the only information is the meta-order itself: in the case of public information, one would simply expect jumps in the price). In particular, if the distribution of meta-orders is power-law with exponent $\gamma$ so that
\[
\PP[l> L] \sim \frac{1}{L^\gamma},
\]
the autocorrelation function of order flow decays as
\[
\rho(\tau) \sim \frac{1}{\tau^{\gamma-1}}.
\]

\item The noise trader sends totally random market orders. Recall that these two agents may not correspond to actual agents, but only to behaviors: each agent could contribute to informed and noise trading at the same time.

\item The reaction of the market to these orders is done by the rest of the market, which we will call market makers. Market makers send limit orders and set the bid and ask, and the corresponding volumes. By assuming perfect competition between market makers, we can assume that there is only one market maker. This agent is also virtual and could represent actual market makers as well as arbitragers or any other market participant (so there is no reason to consider any cash limit in the aggregate).

\end{itemize}

A key to our approach is that we assume perfect information, i.e. that the market maker is able to detect the meta-order that is sent by the informed trader. This amounts to saying that he is able to detect the trends since we reason in the aggregate at that point, which does not seem unreasonable.

\begin{remark}
With what we call perfect information, the price does not respond to noise trading: the market maker knows whether the trader is informed or uninformed and sets the price in function of the information uniquely - whereas in the reality he only knows that afterwards, since he needs a signal/noise decomposition. The price path we derive should thus be viewed as be the expectation of the actual price path. The impact of imperfect information will be studied later in the article.

\end{remark}

\section{Market making strategy and bare market impact}

We now find a competitive market making strategy for the market model defined above (i.e. in which market makers cannot make profits on average). A straightforward argument leads us to conclude that the absence of correlations between the market maker's decisions and the informed trades always leads to super-diffusion on long time scales in the presence of long-ranged correlations; to ensure price diffusion as in \cite{bouchaud2}, the market maker's adaptive reaction to correlated market order flow must be to submit correlated limit orders. Following our previous reasoning, we assume that the market maker decomposes his orders into two parts: the orders in response to noise traders and the orders in response to informed traders. The strategy adopted to satisfy the noise traders is simply to refill the queue when a trade occurs, since no information is conveyed so the state of the order book has no reason to change.  In contrast to traditional treatments such as \cite{glosten} or\cite{kyle}, we do not require any notion of fair value nor is there any implication that traders may have information about the ex-post stock price.\\

As we just said, under our assumption of perfect information, noise traders do not change the net volume at the best quote. Therefore, to study transient impact,  we need only consider the interaction between the informed trader and the market maker (i.e. the reaction of the market to the information conveyed by the trades).\\

Suppose that the market maker has already sold $L$ shares to the informed trader, moving the ask price to $p>0$ (without loss of generality, we set the starting price to zero). The market maker then knows that the meta-order is of length at least $L$ and must decide how many limit orders to post at price $p$ (in ticks). Imposing perfect competition so that the expected profit of the market maker for this trade is zero (or equivalently to ensure that the price process is a martingale), the price $p$ must be the expectation of the price after the meta-order is completed (this is very similar to \cite{glosten}). He must then post $v_{p}$ shares at $p$ according to the following rule:

\begin{equation}
p = \EE[{p}_{\infty} \,| v=v_{p}\,,l\geq L]
\label{marting}
\end{equation}
where ${p}_{\infty}=\lim_{t \to \infty} p_t$.\\

In this stylized version of our model, only informed trades move the price which is thus a deterministic function of $L$. The final price $p$ and the reversion price $p_\infty$ are then both deterministic. As we commented above, in real markets, because of asymmetry of information, noise trades also move the price since information is not known a priori (introducing noise in the price path): we should then view $p$ and $p_\infty$ as expectations of actual prices. Denote the total length of the meta-order (unknown to the market maker) by $l$ and  the maximum price reached at the end of the meta-order by $p_{max}$.  Then, either $l > L+v_p$ and the meta-order exhausts the available quantity at $p$, causing the price to increase to $p+1$, or $L \leq l \leq L+v_p$, $p=p_{\max}$,  the meta-order completes, and the price reverts to $p_\infty$.  Denote the probability that the meta-order completes by $q(p)$ (which we will write $q$ to simplify the notations).  Then because $p_t$ is a martingale,
\beq
p = q\,p_\infty+(1-q)\,(p+1)
\label{eq:MartingaleCondition}
\eeq
so the decay after completion is
\beq
\Delta(p):=p-p_\infty=\frac{1-q}{q}.
\label{eq:deltap}
\eeq

Now denote the order size required to consume limit orders up to price $p$ by $L_{p}=\sum _{k=0} ^{p} v_{p}$ and the tail distribution function of meta-order sizes by $\tilde F(L)=\PP(l\geq L)$. By definition, $q=\PP(l < L_{p} | l \geq  L_{p-1})$ and $1-q=\PP(l \geq L_p|l\geq L_{p-1})$ so
\bea
\Delta(p)&=&\frac{\PP(l\geq L_p|l\geq L_{p-1}) }{\PP(l < L_{p} | l\geq  L_{p-1})}\nonumber\\
&=&\frac{\tilde F(L_p)}{\tilde F(L_{p-1})-\tilde F(L_p)}.
\label{eq:ReversionCondition}
\eea
Equation \eqref{eq:ReversionCondition} thus relates the distribution of meta-order sizes to the reversion levels.  We emphasize that equation\eqref{eq:ReversionCondition} is valid for any given meta-order size distribution and not just for the power-law case explored below in Section \ref{sec:PowerLaw}.

\begin{remark}\label{rq:PriceReversion}
Note that the price must revert immediately to ${p}_{\infty}$ after completion if the market maker can tell when the meta-order ends. In real markets, such information is partial and is inferred from observations (cf. section \ref{sec:decay}); the price after completion decays over time to $p_{\infty}$. This violation of the zero-profit condition results in information asymmetry which, as we will see in section \ref{sec:fastIT} can be exploited to devise an optimal execution strategy.
\end{remark}

 But we can go further still. Assuming again that there is perfect competition so that the market maker makes zero profit in expectation for any meta-order size of size $L_p$,  we have
 \beq
 L_p\,p_\infty=\sum_{k=0}^{p}k\, v_{k}
\label{eq:FairPricing}
 \eeq
 which then gives
 \bea
\Delta(p) = p-p_\infty=p-\frac{1}{L_{p}}\sum_{k=0}^{p}k\, v_{k}
\label{eq:DeltapFair}
\eea

\begin{remark}
Equations \eqref{eq:MartingaleCondition} and \eqref{eq:FairPricing} are precisely the Martingale Condition and the Fair Pricing condition of \cite{farmer5} respectively.  However, in our model,  both \eqref{eq:MartingaleCondition}  and \eqref{eq:FairPricing}  result from imposing perfect competition between market makers and not any notion of fair pricing.
\end{remark}

\begin{remark}
Whereas making zero profit on average for all order sizes naturally comes from perfect competition, the zero profit condition for each size can be viewed as a second order condition: it is indeed the strategy which minimizes $P\&L$ volatility for the market maker - and makes his profit independent from possible fluctuations or estimation errors.
\end{remark}

\begin{remark}
To be consistent, we should impose condition \eqref{eq:FairPricing} on orders of all possible sizes $l$ and not just $l=L_p$ for some $p$.  This however just complicates the argument and does not change our asymptotic results. A comment should be done here about the tick. Indeed, if one wants to impose the above conditions for any order size $l$, one has to consider a continuous underlying price. Our equations are accurate for $l=L_p$ because then the underlying prices reaches an integer multiple of the tick - then the trade is done at the fair price. The difference between the underlying price and the actual discrete price, of order one tick, is actually the premium associated with queue priority - and is a compensation for putting orders long in advance. To state our condition in all generality for all $l$, we should write that the average price paid is the fair price plus this queue priority premium.
\label{rq:approx1}
\end{remark}

\noindent After a little manipulation, we find that \eqref{eq:DeltapFair} may be rewritten as
\beq
\Delta(p)=\frac{1}{L_{p}}\,\sum_{k=0}^{p-1}L_{k}.
\label{eq:Deltap2}
\eeq
In particular, since $L_k \leq L_p$ for all $k \leq p$, we must have
$
0 \leq \Delta(p) \leq p
$.   It follows that for any $p>0$, $\Delta(p)$ can be expressed in the form
\[
\Delta(p)=\alpha(p)\,p
\]
with $\alpha(p)\in (0,1]$.  That is, the trade price always reverts after completion, no matter what the distribution  of meta-order sizes is.

\subsection{Solving for the latent order book}

We may then compute the $v_p$ recursively by equating \eqref{eq:ReversionCondition} and \eqref{eq:Deltap2}:
\[
\frac{\tilde F(L_{p-1}+v_p)}{\tilde F(L_{p-1})-\tilde F(L_{p-1}+v_p)}=\frac{1}{L_{p-1}+v_p}\,\sum_{k=0}^{p-1}L_{k}.
\]

\noindent Next, we derive a recursive expression for the $\Delta(p)$ as follows.  From \eqref{eq:ReversionCondition},
\[
F(L_{p})=\frac{F(L_{p-1})}{1+\frac{1}{\Delta(p)}}
\]
so
\beq
F(L_{p})=\prod_{k=1}^{p} \left(1+\frac{1}{\Delta(k)}\right)^{-1}\, F(L_{0})
\label{eq:Telescope1}
\eeq
which gives us a recursive relation for $\Delta(p)$:

\bea
\Delta(p)=\sum_{k=0}^{p-1}\frac{F^{-1}\left[ \prod_{i=1}^{k}\left( 1+\frac{1}{\Delta(i)} \right)^{-1}F(L_{0}) \right]}{F^{-1}\left[ \prod_{i=1}^{p}\left( 1+\frac{1}{\Delta(i)} \right)^{-1}F(L_{0}) \right]}
\label{eq:Delta}
\eea

Together, for any given meta-order size distribution $F(\cdot)$, equations \eqref{eq:Telescope1} and \eqref{eq:Delta} allow us to plot the evolution of the stock price during execution and the ultimate price level after completion (corresponding to permanent impact).  We can see that both depend only on $F(\cdot)$, knowledge of which according to the order-splitting model of \cite{farmer5} is equivalent to knowledge of the shape of the autocorrelation function of order signs. The parameter $L_{0}$ (which is also the volume at $p=0$) is however not fixed by these equations; it sets the scale for the market impact function. As we will see in section \ref{sec:C}, it directly depends on the volatility and can be uniquely determined.

\subsection{Price volatility}\label{sec:C}

The scale of the market impact function, which is \textit{a priori} unknown (since $L_{0}$ is not fixed), can be directly related to the price volatility (and can thus easily be estimated). Indeed, for a meta-order of size $l$, if we note $p_{max}(l)$ the maximum prices reached after completion and $p_{\infty}(l)$ the reversion price, the total volatility per unit volume is

\bea
\sigma_{l}^{2}=\frac{p_{\infty}(l)^{2}}{l}=\frac{(p_{max}(l)-\Delta(p_{max}(l)))^{2}}{l}
\eea

Assuming a participation rate $\mu$ for the informed trader, the actual volatility is then

\bea
\sigma^{2}=\mu \frac{\EE\left( p_{max}(l)^{2}(1-\alpha(p_{max}(l)))^{2} \right)}{\EE(l)}
\label{eq:scaleMI}
\eea

\begin{remark}
The participation rate $\mu$ can be estimated from the correlations. Indeed, for a given asymptotic behavior of the correlation function (and thus a given distribution for the meta-orders), the scale is fixed by $\mu$. More details will be given in the next section in the case of power-law correlations.
\end{remark}

\begin{remark}
The volatility we compute here is the long-term volatility, corresponding to the scale where the price is diffusive. But we can also see that the instantaneous volatility clearly depends on the current state of the market and the current correlations - its exact computation would require to add a real reaction noise traders though.
\end{remark}

\section{Power-law distribution of meta-order sizes}\label{sec:PowerLaw}

When meta-order sizes are power-law distributed with exponent $\gamma$ so that $F(x) \sim x^{-\gamma}$,  equation  \eqref{eq:ReversionCondition} simplifies to
\begin{equation}
\Delta(p)=p-p_\infty=\frac{1}{\left(\frac{L_{p}}{L_{p-1}}\right)^{\gamma}-1}=\frac{1}{\left(1+\frac{v_{p}}{L_{p-1}}\right)^{\gamma}-1}
\label{eq:PowerLawReversion}
\end{equation}
which again gives a relation between the volume that the market maker must add when the price reaches $p$ and $L_{p-1}$ market orders have been executed, and the post-completion price ${p}_\infty$, such that the price process is a martingale and expected profit is zero.

\begin{figure}[h]
\begin{center}
\includegraphics[scale=0.5]{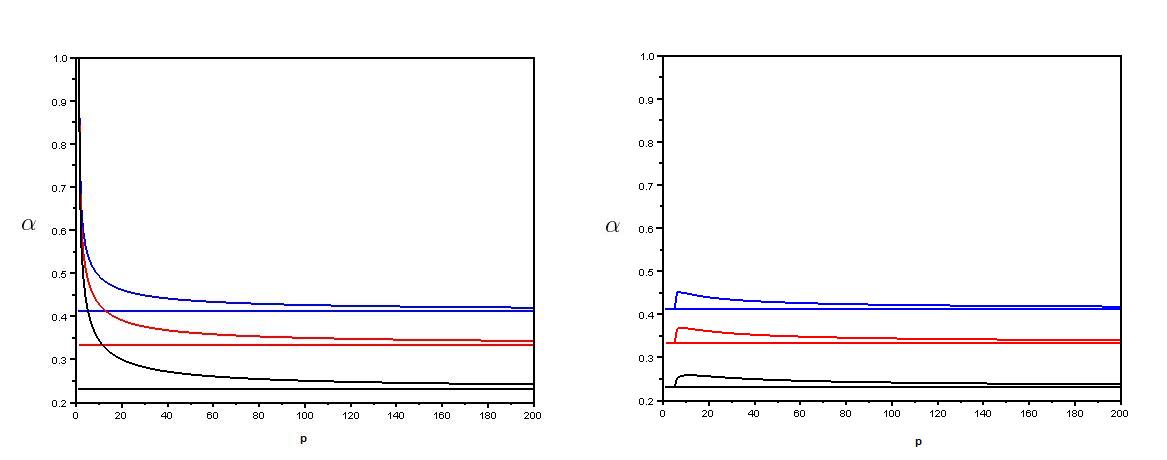}
\caption{$\alpha(p)$ for $F(L)={L^{-\gamma}}$ \textit{(left)} and $F(L)\sim{L^{-\gamma}}$ \textit{(right)}; \textit{black} $\gamma=0.3$, \textit{red} $\gamma=0.5$, \textit{blue} $\gamma=0.7$. As an indication, the corresponding asymptotic values $\alpha_\infty={(\gamma-1)}/{\gamma}$. The singular behavior for small $p$ is due to the approximations described in remarks \ref{rq:approx1} and \ref{rq:approx2}.}
\label{fig:alpha}
\end{center}
\end{figure}

From the definition of $\alpha(p)$ and equation \eqref{eq:Delta}, we obtain the following recursive formula for $\alpha(p)$:
\bea
\alpha(p)=\frac{1}{p}\sum_{k=0}^{p-1}\frac{1}{ \prod_{i=k+1}^{p-1}\left( 1+\frac{1}{\alpha(i)\,i} \right)^{\frac{1}{\gamma}}}
\label{eq:AlphaPowerLaw}
\eea
Plots of $\alpha(p)$ for several values of $\gamma$ are presented in Figure \ref{fig:alpha}.

\begin{remark}
The factors for small $k$ in \eqref{eq:AlphaPowerLaw} are accurate only if $F(L)={L^{-\gamma}}$ for any $L$, whereas of course we only have $F(L)\sim {L^{-\gamma}}$. Thus \eqref{eq:AlphaPowerLaw} should give the correct asymptotic behavior of $\alpha(p)$ for large $p$, but not for small $p$.
\label{rq:approx2}
\end{remark}

If $\alpha(p)$ converges to some $\alpha_{\infty}\in(0,1]$ consistent with the plots in Figure \ref{fig:alpha} we may compute this limit.  Taking  logarithms of \eqref{eq:Telescope1}  gives
\beas
\log L_{p} &=&\log L_{0} + \frac{1}{\gamma}\,\sum _{k=1}^{p}\,\log\left(1+\frac{1}{\alpha(k) \,k}\right)\nonumber\\
&\sim& \frac{\log p}{\alpha_{\infty}\,\gamma}\quad \text { as } p \to \infty
\eeas
so
\begin{equation}
p(L) \sim C\, L^{\alpha_{\infty}\,\gamma} \quad \text { as } L \to \infty.
\label{eq:MarketImpact}
\end{equation}

\noindent The perfect competition condition for the market maker can now be approximated for large $p$ by:

\[
p_{\infty} = \frac{1}{L_{p}}\int _{0}^{L_{p}}\,p(L)\,dL = C\,\frac{1}{L_{p}}\int _{0}^{L_{p}}L^{\alpha_{\infty}\,\gamma}dL = \frac{p}{\alpha_{\infty}\,\gamma+1}
\]
which then gives
\[
1-\alpha_{\infty}=\frac{1}{\alpha_{\infty}\,\gamma+1}
\]
and finally, consistent with Figure \ref{fig:alpha},
\[
\alpha_{\infty}=\frac{\gamma-1}{\gamma}.
\]

\noindent We now have the entire price path, assuming power-law $F(\cdot)$ or equivalently a power-law autocorrelation function of order signs:

\begin{itemize}

\item While the meta-order is being executed, we have from \eqref{eq:MarketImpact} that
\bea
p \sim C L_{p}^{\delta}.
\label{eq:MarketImpact2}
\eea
with $\delta := \gamma-1$.
\item After completion, the price reverts to the level
\bea
p_{\infty}\sim\frac{1}{\gamma}\, p=\frac{1}{1+\delta}\, p.
\label{eq:reversion}
\eea
\end{itemize}

\noindent Square-root impact obviously corresponds to $\delta=1/2$ in which case $(1-\alpha_{\infty})=2/3$ as found in \cite{farmer5}.

\begin{remark}
The relationship \eqref{eq:MarketImpact} between market impact and order size is very reminiscent of the work of Gabaix et al.
\end{remark}

\begin{remark}
Had we performed the above computations with $F$ exponential rather than power-law, we would have found exponential transient impact and a relative permanent impact $(1-\alpha(p))\rightarrow 0$. Our model computations assuming $F(L) \sim L^{-\gamma}$ are consistent with empirical estimates of both the autocorrelation function of order flow (\cite{bouchaud} for example) and the evolution of the price during meta-order execution.
\end{remark}

\begin{remark}
Note that we do not assume that the size distributions  of buy and sell meta-orders are the same. Consequently,  buy and sell orders may have different market impact.
\end{remark}

\begin{figure}[h]
\begin{center}
\includegraphics[scale=0.4]{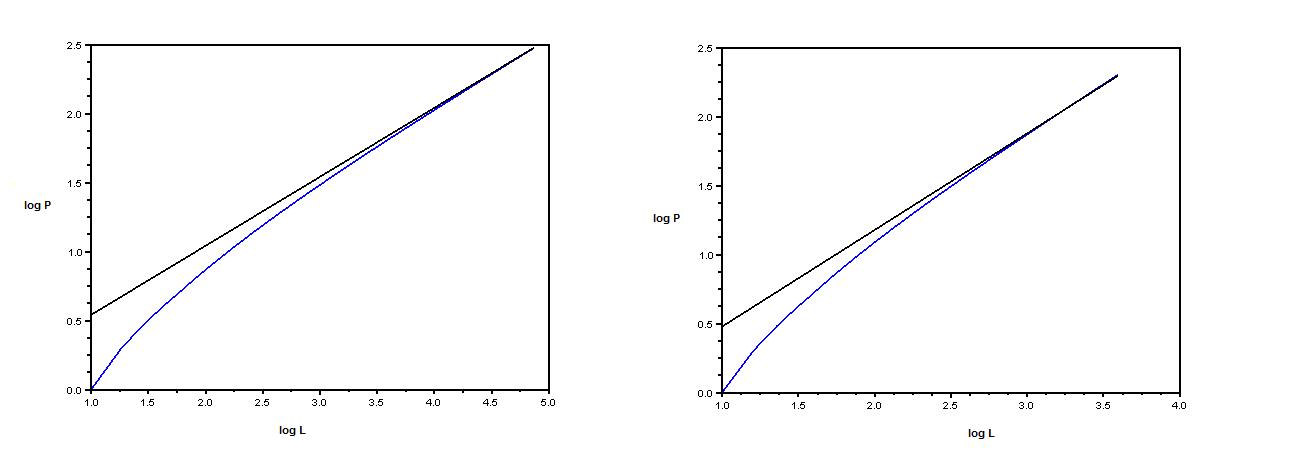}
\caption{The theoretical market impact function, for \textit{(left)} $\alpha=1/3$ and $\gamma=0.5$ and \textit{(right)} $\alpha=1/3$ and $\gamma=0.7$, in log-log scale. For comparison, a line of slope \textit{(left)} 0.5 and \textit{(right)} 0.56.}
\label{1}
\end{center}
\end{figure}

\begin{figure}[h]
\begin{center}
\includegraphics[scale=0.4]{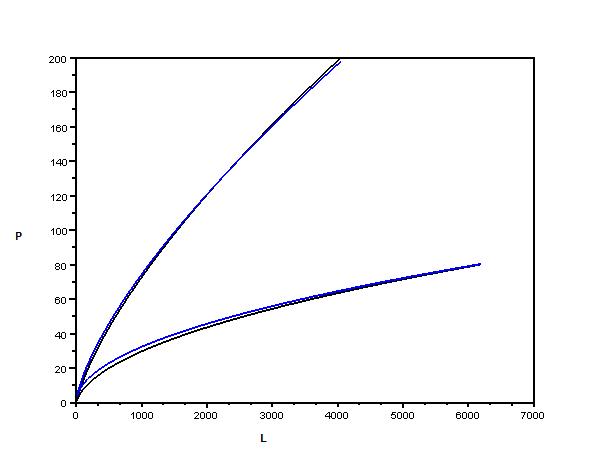}
\caption{The same curves as above, in normal scale. For each couple $(\alpha, \gamma)$, the theoretical impact function is almost equal to the corresponding power-law extrapolation.}
\label{2}
\end{center}
\end{figure}

As we mentioned before, the scale of the market impact function, i.e. the constant $C$, can be related to the volatility per unit volume. If we put together the results \ref{eq:MarketImpact2}, \ref{eq:reversion} and eq.\ref{eq:scaleMI},

\bea
\sigma^{2}=\mu \frac{\EE\left(\left(\frac{1}{\delta+1}\right)^{2} C^{2}l^{2\delta} \right)}{\EE(l)}=\mu \left(\frac{1}{\delta+1}\right)^{2} C^{2} \frac{\EE\left(l^{2\delta}\right)}{\EE(l)}
\eea

which gives the value for $C$:

\bea
C=(\delta+1)\sigma \sqrt{\frac{\EE(l)}{\mu\EE\left(l^{2\delta}\right)}}
\eea

The asymptotical expression of the market impact is therefore

\bea
p_{max}(l)=(\delta+1)\sqrt{\frac{\EE(l)}{\mu\EE\left(l^{2\delta}\right)}}\sigma l^{\delta}
\eea

and the permanent impact is

\bea
p_{\infty}(l)=\sqrt{\frac{\EE(l)}{\mu\EE\left(l^{2\delta}\right)}}\sigma l^{\delta}
\eea

\begin{remark}
The impact is proportional to the volatility, which is what one would expect.
\end{remark}

\begin{remark}
The term $\sqrt{\frac{\EE(l)}{\EE\left(l^{2\delta}\right)}}$ is close to 1 when $\gamma\simeq 1.5$ (and $\delta=\gamma-1=0.5$), which is the case in practice.
\end{remark}

\begin{remark}
Until now we have considered unit volumes. One can also express the market impact in function of the volatility per trade $\sigma_{T}$, by taking into account the average volume of trades $\overline{n}$:

\bea
p_{max}(l)=(\delta+1)\sqrt{\frac{\EE(l)}{\mu\EE\left(l^{2\delta}\right)}}\frac{\sigma_{T}}{\sqrt{\overline{n}}} l^{\delta}
\eea

One could equivalently consider the distribution of $\frac{l}{\overline{n}}$ instead of $l$ in the whole work.
\end{remark}

\begin{remark}
For $\delta=1/2$, $p_{max}(l)=cte .\sigma\sqrt{\frac{l}{\overline{n}}}$, which is exactly the relationship pustulated in \cite{bouchaud}.
\end{remark}

\subsection{Correlations under power-law market impact: what about the reverse?}

We have so far derived the expression of market impact in the case where the sign process gives rise to power-law correlations with a given exponent $\gamma$, without wondering where this exponent comes from. We do not intend to prove that the whole state of the market is optimal, but to show that the hypothesis is consistent with the results, if one considers the inverse problem.\\

Gabaix has shown that for the observed distribution of fund sizes (i.e. power law with exponent 1), and for the behavior described below, the square-root market impact ($\delta=1/2$) leads to the exponent $\gamma=1.5$ for the correlations: this is consistent with our model since we find that $\delta=\gamma-1$. We can even generalize his result for any $\delta>0$. Indeed, if one makes the computation for any market impact exponent $\delta$, one finds that the exponent of the correlation function is $\gamma=\delta+1$. The results are thus consistent, and one can assume that the power-law behavior is a stable equilibrium.

\begin{remark}
The exponent $\gamma$ is not fixed by these results, which means that it can potentially vary across the markets, the countries, etc without losing the optimality (at least to the extent that the markets are uncorrelated). Highly correlated markets are likely to have similar exponents though.
\end{remark}

\subsection{Empirical estimation of $\delta$}

\begin{figure}[h]
\begin{center}
\includegraphics[scale=0.5]{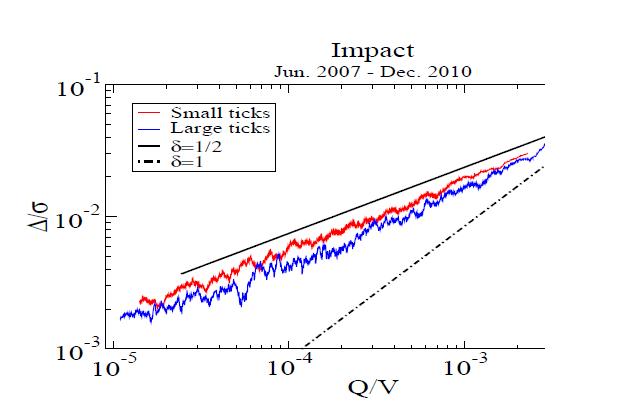}
\caption{Market impact for Capital Fund Management proprietary trades on futures markets, in the period June 2007 - December 2010 (\cite{bouchaud}).}
\label{MIjpb}
\end{center}
\end{figure}

The well-known square-root formula for market impact is consistent with price evolution during meta-order execution of the form $p \sim \sqrt{L}$.  On the other hand, various empirical studies indicate that $p \sim L^\delta$ with $\delta \in (0.5,0.7)$ (fig.\ref{MIjpb}).  Either way, in our model, there is a one-to-one correspondence between the average evolution of the price during meta-order execution and the distribution of meta-order sizes: the exponent of the market impact function can be estimated by measuring the correlations in the order flow.

\subsection{The informational content of trades and diffusion}

In our model, every (informed) trade has permanent impact, which means that information is indeed incorporated in the price consistent with economic theory.  Yet this is not in contradiction with price diffusion, for the simple reason that the impact is not constant but depends on the size of the meta-order. Which, in other terms, means that the information per trade in a large meta-order is smaller. The explanation is simple: the immediate impact of a trade is the inverse of the volume at the opposite best quote, which increases with the size of the meta-order. Therefore, our model not only considers the impact of market orders, but a joint impact of market and limit orders: this is a way to include the order book shape and the order flow history in the computation of market impact.

\subsection{The latent order book}

\begin{figure}[h]
\begin{center}
\includegraphics[scale=0.45]{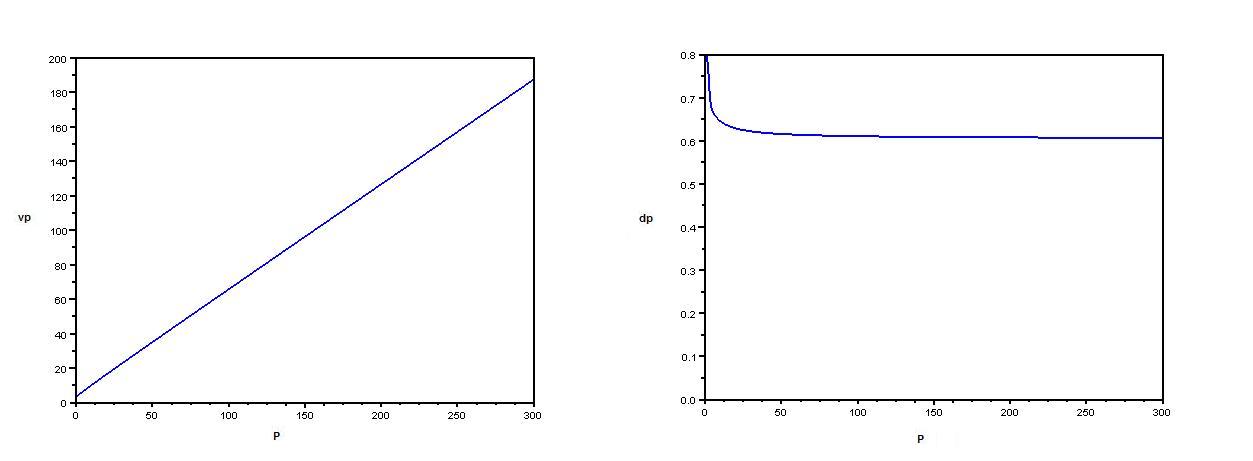}
\caption{\textit{(left)} The latent volume $v_{p}$ for $\gamma=0.5$ and thus $\alpha=1/3$. \textit{(right)} The marginal volume $d_{p}$.}
\label{fig:latent1}
\end{center}
\end{figure}

\begin{figure}[h]
\begin{center}
\includegraphics[scale=0.45]{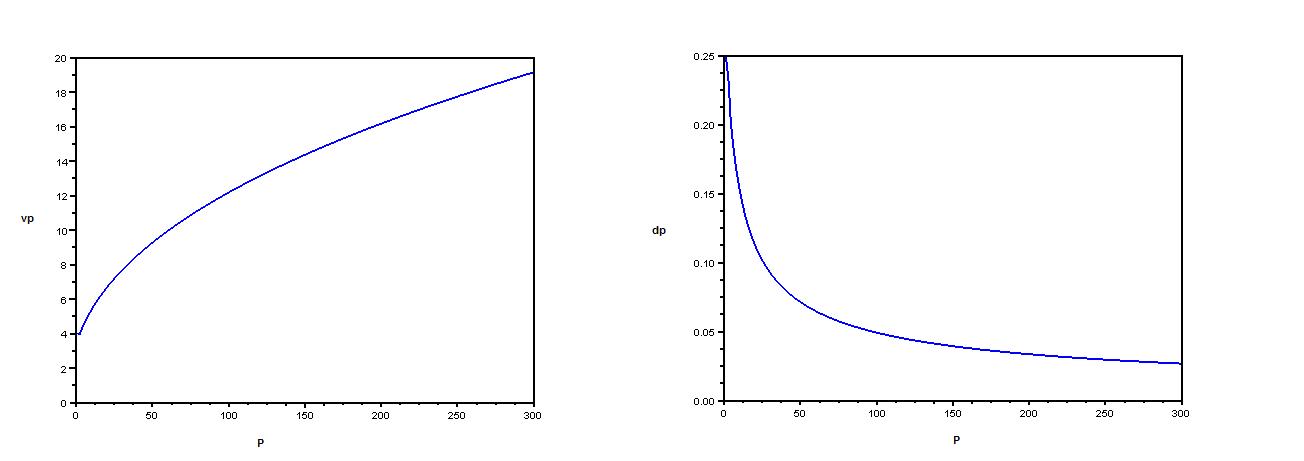}
\caption{\textit{(left)} The latent volume $v_{p}$ for $\gamma=0.7$ and thus $\alpha\simeq 0.41$. \textit{(right)} The marginal volume $d_{p}$. The latent volume is now sub-linear.}
\label{fig:latent2}
\end{center}
\end{figure}

With $\delta \approx 1/2$ as empirically observed, our model leads naturally to a quasi-linear latent order book as defined in \cite{bouchaud} (exactly linear if $\delta=\gamma-1 =1/2$). Specifically, the volume $v_{p}$ that the market maker posts at price $p$ can be interpreted as the latent volume that would emerge if the price were to reach $p$. Our model provides an explanation for the quasi-linearity of the latent order book postulated in \cite{bouchaud}: In our model,  this latent order book shape reflects the adaptive reaction of the market-maker under perfect competition to autocorrelated order flow.
Fig.\ref{fig:latent1} and \ref{fig:latent2} represent the volume $v_{p}$ in function of $p$ and the increase in volume from price $p$ to price $p+1$: $d_{p}=v_{p+1}-v_{p}$. The expression of the latent volume can be derived:

\bea
v_{p}=L_{p}-L_{p-1}&\sim& \left( \frac{p}{C}\right)^{\frac{1}{\gamma-1}}-\left( \frac{p-1}{C}\right)^{\frac{1}{\gamma-1}}\nonumber\\
&\sim&\left( \frac{p}{C}\right)^{\frac{1}{\gamma-1}}\left(1-\left( 1-\frac{1}{p}\right)^{\frac{1}{\gamma-1}} \right)\nonumber\\
&\sim&\frac{1}{\gamma-1}\left( \frac{p}{C}\right)^{\frac{1}{\gamma-1}-1}
\eea

For $\gamma=1.5$, we find a linear latent volume.

\subsection{Diffusivity of the price}

\begin{figure}[h]
\begin{center}
\includegraphics[width=\linewidth]{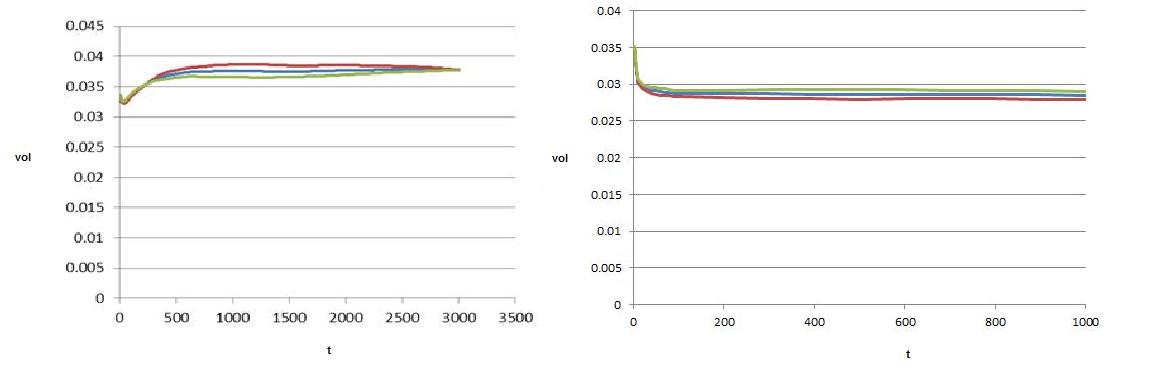}
\caption{Volatility as a function of time for $\gamma=0.7$, for \textit{(left)} the  $\epsilon$-Intelligence Market for a non-zero cancelation rate and \textit{(right)} our model. We see that our model ensures diffusion even though market order flow is highly autocorrelated. }
\label{fig:Diffusion}
\end{center}
\end{figure}

It is straightforward to verify by simulation that the price diffuses on almost all timescales in our model as we ensured by imposing the Martingale Condition \eqref{eq:MartingaleCondition}.  Philosophically, the market price diffuses in our model because autocorrelation in the flow of market orders which would otherwise result in superdiffusion of the price,  is compensated by autocorrelation in the limit order flow corresponding to the adaptive reaction of the market maker under perfect competition, as empirically observed in \cite{eisler}. If limit orders are only passive and thus are not correlated with market orders, one can show that the price cannot diffuse for long time scales, as illustrated on figure \ref{fig:Diffusion}.

To be more precise, one can show that if limit orders are passive the volatility per trade behaves like $\sigma^{2}_{t}\sim t^{2-\gamma}\left(1-a\frac{t^{\gamma}}{T}\right)$ where $t$ is the period on which the volatility is computed, $T$ the total period that is considered and $0<a<1$ is a constant (this asymptotic formula is accurate provided $t^{\gamma}<T$ ).The reasons for this behavior are the following:

\begin{itemize}
\item for $t\ll T^{\frac{1}{\gamma}}$, the growth proportional to $t^{2-\gamma}$  is due to the power-law correlations in market orders.
\item but the events that are too rare to occur in the period $T$ are not observed with high probability. Thus, instead of increasing as a power-law, the volatility stagnates with high probability at a level proportional to $T^{\frac{2-\gamma}{\gamma}}$. Of course, this stagnation is artificial and cannot be seen as diffusion.
\end{itemize}

These two regimes are indeed observed on fig \ref{fig:Diffusion}.

\begin{remark}
We recall that prices are diffusive when $\sigma^{2}_{t}$ does not depend on $t$.
\end{remark}

\subsection{Model for the decay of market impact}\label{sec:decay}

As we have seen before, the decay of market impact after completion of the meta-order is a key of the competition between the market maker and the informed trader since it allows the informed trader to execute his order at a better average price for the same permanent impact (which results in a loss for the market maker). In the continuity of the previous sections, and under the same hypothesis, we derive an expression for the shape of the decay. Here, market makers are no longer assumed to make zero-profit (otherwise there would be no optimal execution strategy for the informed trader), but they are supposed to minimize their loss (and they still cannot make profits on average). Actually the real reason for market makers to be present is the noise trader, from which he always makes profits. The presence of informed traders cannot be profitable for him (otherwise they would not trade), but he can still optimize his strategy as far as they are concerned. A remark is that the fair price condition introduced in \cite{farmer5} is violated under these assumptions. This is a major difference between the two models, since in ours the perfect competition takes place between the market makers, but the informed trader is not assumed to trade at a fair price. In other words, for the same permanent impact (since prices are not manipulable) he can take advantage of the asymmetry of information to try to optimize his execution.\\

Assume that the informed trader was trading at rate $\mu$ until time $t_{max}$ (the price is then $p_{\max}$, the number of informed trades is $L$). While he does not trade again, there are two options (each of one results in a loss for the market maker). Either the informed trader has not finished, in which case the price should remain $p_{\max}$ (if not the loss is $\Delta_{1}(p)$), or he is, and the price should be $p_{\infty}$ (if not the loss is $\Delta_{2}(p)$). Our goal is to quantify these losses and define the price so that at each time the expectation of the loss is minimized. Right after the last trade, there is a large probability that the informed trader is still here, and the price must be close to $p_{\max}$ since $\Delta_{1}(p)$ must be minimized, but each noise trade decreases this probability and after a while the price should return to $p_{\infty}$ (since then $\Delta_{1}(p)$ must be minimized).\\

\begin{figure}[h]
\begin{center}
\includegraphics[scale=0.6]{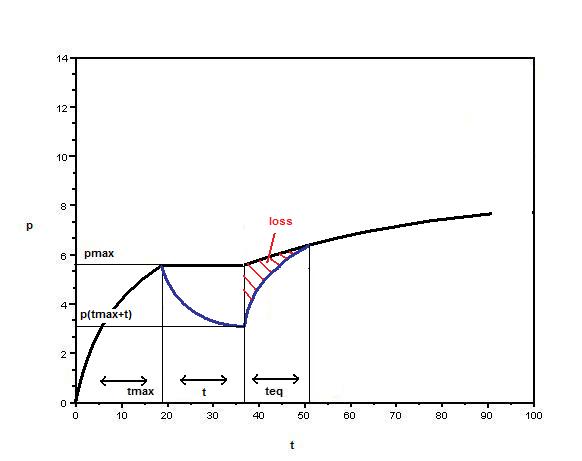}
\caption{Decay of market impact and notations. In blue, the market impact, and in black, the virtual market impact if there were no decay between $t_{max}$ and $t_{max}+t$. The loss $\Delta_{1}(p)$ is represented in red. After $t_{eq}$, the price is back to the price that would be set without decay.}
\label{param}
\end{center}
\end{figure}

Let us compute the loss $\Delta_{1}(p)$ conceded to the informed trader if he resumes trading at time $t_{max}+t>t_{max}$ (loss which is due to the decay: the asset is sold at a lower price), after $L$ trades. Let us define $L+L_{eq}$ as the number of trades after which the price has reached the correct price again (cf. Fig.\ref{param}, at time $t_{max}+t+t_{eq}$), and $I(l)$ the transient market impact function for $l$ informed trades.

The loss conceded to the informed trader during the period if he is still here, in comparison with the case when the informed trader does not stop trading (or the price does not decays) can be approximated by:

\bea
\Delta_{1}(p_{t_{max}+t}) = \int _{0}^{L_{eq}} I(L+l) dl - \int _{0}^{L_{eq}} (p_{t_{max}+t}+I(l))dl
\label{eq:loss1}
\eea

We make here the approximation that the meta-order is still large enough to reach $L+L_{eq}$ (but anyway it would certainly be suboptimal for the informed trader not to trade the maximum quantity below the fair price, and if not it would be arbitraged). If the informed trader is no longer here, the potential loss is:

\bea
\Delta_{2}(p_{t_{max}+t}) = L p_{t_{max}+t}-\int _{0}^{L} I(l) dl
\label{eq:loss2}
\eea

 and is due to the fact that the previous meta-order has been executed on average below $p_{t_{max}+t}$, which is the new reference price for the new possible meta-order. In assuming that the market maker can detect the end of a meta-order, one would only consider $\Delta_{2}(p_{t_{max}+t})$ and not $\Delta_{1}(p_{t_{max}+t})$, and find an immediate decay to the fair price (which we will call $p_{fair}$) as in \cite{farmer5}. $p_{t_{max}+t}$ must be fixed according to:

\begin{equation}
p_{t_{max}+t}=arg\min _{p_{fair}\leq p \leq p_{max}} (P(1|t)\Delta_{1}(p)+P(2|t)\Delta_{2}(p))
\end{equation}

where $P(1|t)$ and $P(2|t)$ are the probabilities of the different events given that $t$ noise trades have occurred since the last informed trade. We bound $p$ since because of the competitive behavior the market maker cannot make money on average, either by increasing the price over $p_{max}$ or by decreasing it below $p_{\infty}$. In addition, the following relation defines $L_{eq}$:

\begin{equation}
p_{t_{max}+t}+I(L_{eq})=I(L+L_{eq})
\end{equation}

By injecting this expression in the previous minimization, we find:

\begin{eqnarray*}
L_{eq}= arg\min _{p_{fair}\leq I(L')\leq p_{max}}( P(1|t)[\int _{0}^{L'} I(L+l) dl - \int _{0}^{L'} (I(L+L')-I(L')+I(l))dl] \\
+ P(2|t)[L (I(L+L')-I(L'))-\int _{0}^{L} I(l) dl ])
\end{eqnarray*}
\begin{equation}
\end{equation}

By taking the derivative, we find the very simple condition (without the condition $p_{fair}\leq I(l')\leq p_{max}$):

\begin{equation}
\frac{L_{eq}}{L}=\frac{P(2|t)}{P(1|t)}
\end{equation}

And thus the value of $p_{t_{max}+t}$:

\begin{equation}
p_{t_{max}+t}=max(\frac{1}{L}\int _{0}^{L} I(l) dl, I(\frac{L}{P(1|t)})-I(\frac{L P(2|t)}{P(1|t)}))
\end{equation}

Where the first argument is $p_{fair}$. Before using the explicit forms of the quantities above to find the decay function, let us make some comments about the general form. If $P(1|t)=1$, which means that the informed trader is still here with probability 1, this formula gives $p_{t_{max}+t}=I(L)$, which means that $p_{t_{max}+t}=p_{max}$ (which is what we would expect). But if on the contrary $P(1|t)=0$, then $p_{t_{max}+t}=p_{fair}$ since the second argument tends to 0.\\

For an impact function of the form $p=CL^{\gamma}$,

\begin{equation}
\frac{p_{t_{max}+t}}{p_{max}}=max(\frac{1}{\gamma+1}, (\frac{1}{P(1|t)})^{\gamma}-(\frac{P(2|t)}{P(1|t)})^{\gamma})
\end{equation}

We notice that this gives directly the ratio $\frac{p_{t_{max}+t}}{p_{max}}$ : the process is scale invariant with respect to $p$. Moreover, it would not seem unrealistic to assume that the market maker estimates $P(1,t)$ as a function $f(t/t_{max})$ (this would mean that he considers the trading time as a characteristic time of the informed trader). This would give a scale invariance with respect to $t$. For example, assume that

\begin{equation}
P(1|t) = (1-\frac{1}{tmax})^{t} \simeq e^{-\frac{t}{t_{max}}}
\end{equation}

then

\begin{equation}
\frac{p_{t_{max}+t}}{p_{max}}=max(\frac{1}{\gamma+1}, (\frac{1}{(1-\frac{1}{tmax})^{t}})^{\gamma}-(\frac{1-(1-\frac{1}{tmax})^{t}}{(1-\frac{1}{tmax})^{t}})^{\gamma})
\end{equation}

In particular, for $t \ll t_{max}$,

\begin{equation}
\frac{p_{max}-p_{t_{max}+t}}{p_{max}}\sim (\frac{t}{t_{max}})^{\gamma}
\end{equation}

\begin{figure}[h]
\begin{center}
\includegraphics[scale=0.5]{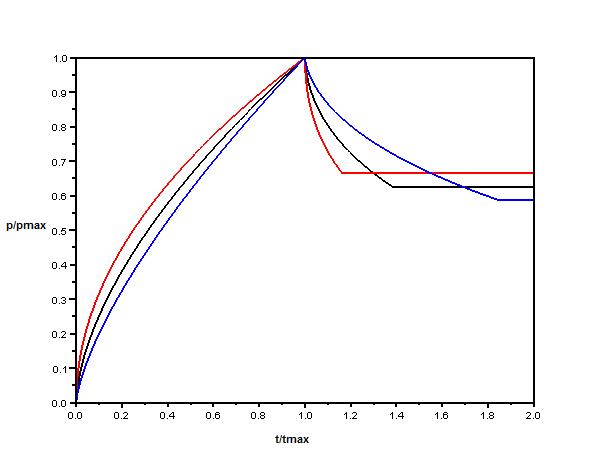}
\caption{The market impact function for (red) $\gamma=0.5$, (black) $\gamma=0.6$ and (blue) $\gamma=0.7$, for the estimator of $P(1|t)$ above. The length of the meta-order and the participation rate do not influence the shape in relative coordinates.}
\label{MItotal}
\end{center}
\end{figure}

Note that $p_{t_{max}+t}$ has been computed for continuous prices. For discrete prices, the market maker would need to minimize on a discrete ensemble (but this would not change any price by more than a tick).\\

Even though the exact shape of the decay may vary according to the estimator, according to our model the impact function comes back to $p_{\infty}$ after a time which is proportional to the trading time $t_{max}$. This is enough information to find optimal execution strategies (cf. section \ref{sec:OptimalExecution}).

\section{Renormalized market impact}\label{sec:MIindiv}

In some way, the previous results are only preliminary and remain abstract. They only allow to define the fundamental properties of the market in which the agents trade, and give his properties in the aggregate. Now, in this framework, the real quantities can be computed (with little effort): in particular the renormalized market impact which is so important to traders.\\

Contrary to what is often done in the literature, we do not assume that each agent has a well-defined impact, and that their sum gives the price process. On the contrary, here the sum of all the agents have a fixed impact, and the impact of a trader depends on the way he fits into this global process - the impact being thus very revealing of the quality of his strategy as we will discuss.

\subsection{The impact on prices}

So far, we have considered only the very stylized scenario where there is only one informed trader and the market maker knows whether his counterparty is informed or uninformed. We now consider the more realistic scenario where an informed trader (trader A) sends market orders while other informed traders are trading.\\

To study the impact of trader A, we must separate his meta-order from the aggregated orders of the other actors (trader B for short). When trader A is not present, trader B behaves as in the previous model, submitting meta-orders thereby generating power-law autocorrelated order flow. We now compute the market impact of a meta-order of length $l$ submitted by trader A while trader B is trading.\\

We assume the following:
\bi
\item Trader A's participation rate $\widetilde{\mu}$ is constant,
\item This participation rate is smaller than the participation rate $\mu$ of trader B: $\widetilde{\mu} \ll \mu $ (which is not necessary but gives simpler results),
\item We call $l_{A}$ the volume that trader A wants to trade.
\ei

If trader A trades a volume $\Delta l$ when both actors have the same sign (resp. an opposite sign), his market impact for this partial volume (provided $\Delta l \ll l$, which will be verified since $\widetilde{\mu} \ll \mu$) is

\bea
p^{A}_{+,l}(\Delta l)=I(l+\Delta l)-I(l)\simeq \Delta l I'(l)\\
p^{A}_{-,l}(\Delta l)=I(l)-I(l-\Delta l) \simeq \Delta l I'(l)
\eea

Under our hypothesis, the total trading time of trader A is of order $\frac{l_{A}}{\widetilde{\mu}}$. His total permanent impact (which is now the only relevant quantity, since the trades are quite spaced) is the sum of his impacts in each period where trader B has a defined trend (i.e. $p^{A}_{+,l_{i}}(\Delta l_{i})$ if trader B buys a volume $l_{i}$ in his $i^{th}$ period, $p^{A}_{-,l_{i}}(\Delta l_{i})$ if he sells it). $\Delta l_{i}$ is here the volume that trader A buys in the $i^{th}$ period.\\

Now if the trading rate of trader A is constant, if we define $N=\min \left( n | \sum_{j=0}^{N}\frac{l_{j}}{\mu}>\frac{l_{A}}{\widetilde{\mu}} \right)$ as the number of times that the trend of trader B changes during the trade (and re-defining $l_{N}$ so that $\sum_{j=0}^{N}\frac{l_{j}}{\mu}=\frac{l_{A}}{\widetilde{\mu}}$),

\bea
\Delta l_{i}=\frac{l_{i}}{\sum_{j=0}^{N}l_{j}}l_{A}
\label{eq:li}
\eea

Therefore, knowing the lengths $l_{i}$ and therefore $N$ the total impact is:

\bea
I(l_{A})&=&\sum_{i=0}^{N} \Delta l_{i} I'(l_{i})\nonumber\\
&=&\sum_{i=0}^{N} \frac{l_{i}I'(l_{i})}{\sum_{j=0}^{N}l_{j}}l_{A}\nonumber\\
&=&\frac{\widetilde{\mu}}{\mu}\sum_{i=0}^{N}l_{i}I'(l_{i})
\eea

The average renormalized impact function $I(l_{A})$ is therefore:

\bea
\overline{I}(l_{A})=\frac{\widetilde{\mu}}{\mu} \EE\left(\sum_{i=0}^{N}l_{i}I'(l_{i}) | \sum_{j=0}^{N}\frac{l_{j}}{\mu}=\frac{l_{A}}{\widetilde{\mu}} \right)
\eea

In the case where correlations are power-law, this becomes

\bea
\overline{I}(l_{A})=\frac{\widetilde{\mu}}{\mu} \frac{C\delta}{\delta+1}\EE\left(\sum_{i=0}^{N}l_{i}^{\delta} | \sum_{j=0}^{N}l_{j}=\frac{\mu}{\widetilde{\mu}}l_{A} \right)
\eea

\begin{figure}[h]
\begin{center}
\includegraphics[scale=0.4]{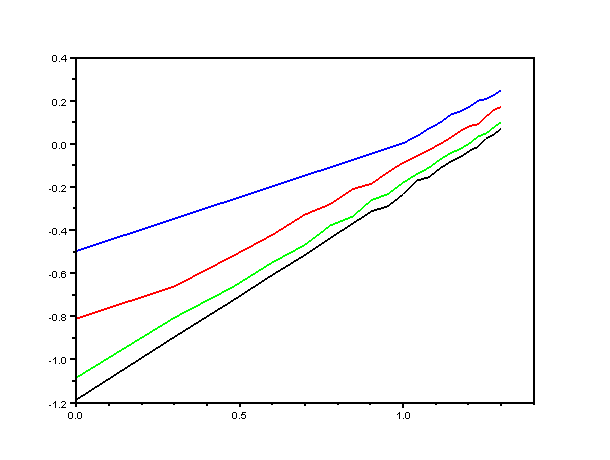}
\caption{The average renormalized market impact function $I(l_{A})$ versus the volume traded by trader A (in log-log scale), with \textit{(blue)} $\widetilde{\mu}=0.5\mu$, \textit{(red)} $\widetilde{\mu}=0.1\mu$, \textit{(green)} $\widetilde{\mu}=0.01\mu$ and \textit{(black)} $\widetilde{\mu}=0.001\mu$. The renormalized impact tend from power-law to linear when the trading rate decreases. Note that the slower the trading takes place, the lesser the market impact.}
\label{fig:MIlong}
\end{center}
\end{figure}

\begin{remark}
In the particular case where there is only one period (either because $l_{A}$ is very small or because $l_{0}$ happens to be very large), $\overline{I}(l_{A})\sim l_{A}^{\delta}$. The constant is smaller than $\frac{C}{\delta+1}$ though because the trades of trader A are mixed with the trades of trader B. The opposite situation (ie for large $N$) is studied below, and in this case the market impact is linear. The reality is between these limit case. If the trading time of trader A is small, his market impact is square root. But when he trades slower, it tends to be linear (cf. fig.\ref{fig:MIlong}).
\end{remark}

If we don't take into account the side effect due to $l_{0}$ and $l_{N}$ (which becomes true for large $N$),
\bea\overline{I}(l_{A})=\frac{\widetilde{\mu}}{\mu} \frac{C\delta}{\delta+1} \frac{\EE\left(l_{0}^{\delta}\right)}{\EE\left(l_{0}\right)}l_{A}
\eea
which means that the asymptotic impact is linear.

\begin{remark}
The market impact always tends to be linear on long time scales, which rules out price manipulation on that time scales.
\end{remark}

\begin{remark}
This linearity rules out the possibility of a zero permanent impact. Indeed for each new period the impact of the trades submitted in the previous periods would decrease to 0 and would not add so as to obtain the linearity.
\end{remark}

\begin{remark}
Contrary to the measurement using the order flow imbalance (cf. section \ref{sec:OFI}), the finiteness of the set does not affect the exponent, but introduces a multiplicative constant of the form $(1-\frac{a}{l_{max}})$ so that the empirical findings always under-estimate the bare impact (the constant tends to 1 for an infinite set though).
\end{remark}

\begin{remark}
Throughout the end of the paper, the market impact function will be noted $Cl^{\delta}$, where $C$ takes into account the appropriate rescaling effects.
\end{remark}

\begin{remark}
We see a difference appear between the bare market impact (for market trends) and the renormalized impact of individual trades (which can be executed over several trends). This difference explains the apparent paradoxical co-existence of absence of manipulation and concave impact.
\end{remark}

\subsection{The observed impact}

As we discussed above, the observed impact may vary a lot among the actors.  For example, If an actor is always a trend-leader, he will observe a smaller decay after his trade than an actor who is always a trend-follower (since one is at the beginning of the global trend and the other at the end, and the reversion price for the global trend is the same for both).\\

The observed transient impact may also vary, leading to various power-law exponents if one tries to make estimations (\cite{moro}). We have studied the case of a VWAP (constant trading rate), in which case the strategy is independent from the background trends, and the impact (observed and effective) is power-law with the same exponent than the bare impact and becomes linear on long time scales. But trading strategies with varying rate may lead to different observed impacts - different exponents for instance. One could even assume that some traders are on average anti-correlated with the market, so that their observed impact is negative.

\section{Price change and order flow imbalance}\label{sec:OFI}

In our analysis, the impact is computed by someone who knows when the meta-order - or the trend - begins. This can be computed by trading firms that know exactly when their order begins and ends (or anyone provided one has some information about the origin of every trade). If one wants to compute the market impact without knowing from whom the trades come in estimating the price changes in function of the global imbalance in order flow, the story is quite different.\\

Assume that one observes the impact of a global imbalance $Q$ for $L$ consecutive trades randomly chosen, by estimating the difference between the price of the first trade and the price of the last trade. We will call this the \textit{Aggregate Market Impact} If we note $N$ the number of trends during the period, $l_{i}$ the volume associated to each trend and $I(l_{i})$ the permanent market impact associated to these trades, then

\bea
I_{aggr}(Q,L)=\EE  \sum_{i=1}^{N} \epsilon_{i}I(l_{i})-\epsilon_{1}I(l_{1,<}) - \epsilon_{N}(I(l_{N})-I(l_{N,>}))  | &\sum_{i=1}^{N}& \epsilon_{i}l_{i}-\epsilon_{1}l_{1,<}-\epsilon_{N}(l_{N}-l_{N,>})=Q, \nonumber\\
&\sum_{i=1}^{N}& l_{i}-l_{1,<}-(l_{N}-l_{N,>})=L
\label{eq:aggregatedMI}
\eea

where $l_{1,<}$ is the volume belonging to the first trend which was traded before the beginning of the observation and $l_{N}-l_{N,>}$ is the volume belonging to the last trend which is traded after the end of the observation.

\begin{remark}
To be exact, one should consider the transient impact for the last period, since for this period the price has not reverted yet to its permanent level. We do not consider this side effect in order to simplify the computations.
\end{remark}

For $N=1$, $L=Q$ and

\bea
I_{aggr}(Q,1)&=&\EE\left[ I(l_{1,<}+L)-I(l_{1,<}) | L \right]\nonumber\\
&=&\frac{\sum_{l\geq L} \PP\left( l_{1}=l\right)\sum_{k=0}^{l-L}\left[ I(k+L)-I(k)\right]}{\sum_{l\geq L} \PP\left( l_{1}=l\right)(l+1-L)}\nonumber\\
&\simeq& cte \frac{\sum_{l\geq L} l^{-\delta-2}\sum_{k=0}^{l-L}\left[ (k+L)^{\delta}-k^{\delta}\right]}{\sum_{l\geq L} l^{-\delta-2}(l+1-L)} \nonumber\\
&\simeq& cte \frac{\sum_{l\geq L} l^{-\delta-2} L l^{\delta}}{\sum_{l\geq L} l^{-\delta-2}(l+1-L)} \quad \text{for small $L$} \nonumber\\
&\simeq& cte \frac{\sum_{l\geq L} l^{-2} L}{\sum_{l\geq L} l^{-\delta}}\nonumber\\
&\simeq& cte L^{\delta}
\eea

\begin{figure}[h]
\begin{center}
\includegraphics[scale=0.8]{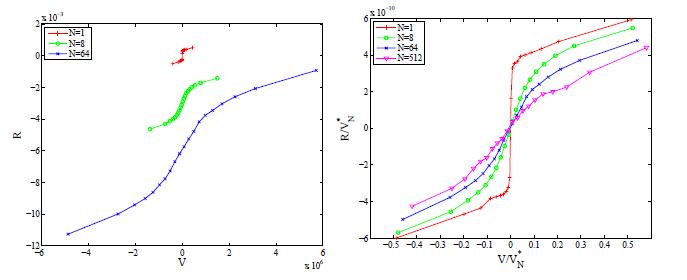}
\caption{Aggregate market impact for the LSE stock Astrazeneca for 2000-2002 (\cite{BFL}).}
\label{MIlargeTS(1)}
\end{center}
\end{figure}

For small $L$ and $N=1$ (which is the case in fig.\ref{MIlargeTS(1)} for the red curve, since it considers only one trade.), the aggregated impact behaves as $L^{\delta}=Q^{\delta}$.

\begin{figure}[h]
\begin{center}
\includegraphics[width=8cm]{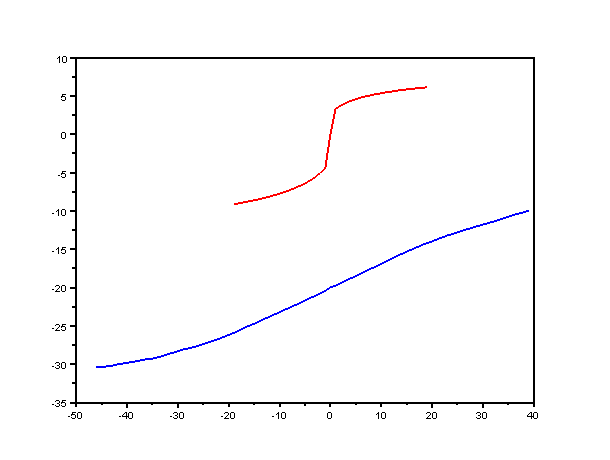}\hfill
\includegraphics[width=8cm]{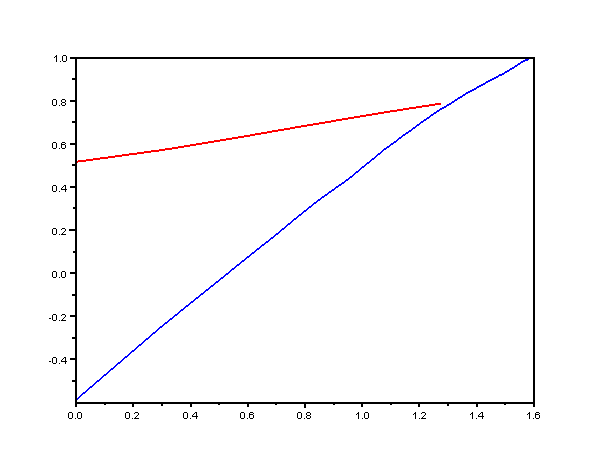}
\caption{Aggregate market impact for different values of $N$, for $\delta=0.5$: strongly concave for $N=1$, then progressively linear for larger $N$. \textit{(left)} In red, the market impact for one observed trade, in function of its volume, in blue, the market impact for several trades, in function of the aggregated order-flow (an intercept of -20 has been added to aid visualization). \textit{(right)} The same plots in log-log scale. The exponent of the apparent impact for single trades is close to 0.25, quite smaller than $\delta$. This is found empirically on fig.\ref{MIlargeTS(1)}}\label{fig:aggregMI}
\end{center}
\end{figure}

\begin{remark}
In the reality, the estimation of $I_{aggr}(Q,L)$ is made for a finite set of events: one do not observe events with small probabilities, i.e. $l>l_{max}$. Introducing this cut-off in the expression of $I_{aggr}(Q,1)$ gives an expression of the form:

\bea
I_{aggr}(Q,1)\simeq cte L^{\delta}(1-cte'\frac{L}{l_{max}})
\eea

which gives an even more concave function than $L^{\delta}$: this is illustrated in fig.\ref{fig:aggregMI}, and observed empirically (cf. fig.\ref{MIlargeTS(1)}). The apparent exponent is therefore smaller than $\delta$ in this particular case.

\end{remark}

On the contrary, in the extreme case where $l_{i}=l_{0}$ $\forall i$),

\bea
I_{aggr}(Q,N-1)=cte Q
\eea

and the impact is linear.

\begin{remark}
One would expect that the exact solution of eq.\ref{eq:aggregatedMI} is roughly linear for large $Q$, and tends to a power-law with exponent of order $\delta$ for small $L$ (and thus small $Q$). The exact solutions of eq.\ref{eq:aggregatedMI} are plotted on fig.\ref{fig:aggregMI}.
\end{remark}

\begin{remark}
The interpretation is that for one (resp. a few) trades, the whole (resp. most of the) volume belongs to the same trend. Thus it is possible to observe a concavity which comes from the concavity of the bare market impact (but there is a priori no reason that both are equal, and even no reason that the exponent is the same as shown in fig.\ref{fig:aggregMI}). On the contrary, when one considers a large number of trades, the market impact is overwhelmed by trend changes and the aggregated impact tends to be linear since the trends are independent. This exactly explains the observations in \cite{BFL} (Fig.\ref{MIlargeTS(1)}).
\end{remark}

\section{Market impact and asymmetry of information, or the impact of arbitragers}

Until now, we have assumed that the market maker was able to detect the informed trades. Under this assumption, and assuming perfect competition between the market makers, we have found an expression for the transient and permanent market impact. We could wonder: what if we introduce asymmetry of information, and the market maker can only estimate the activity of informed traders to define his strategy? Could we take it into account in our study?\\

To take it into account, we would have to upgrade the intelligence of the market maker. This would need assumptions on their behavior, on their estimators, on the different models that they actually use. Not only would that be purely hypothetical, but it would be far from reality. Hundreds of traders have worked for decades to arbitrage the market and contribute in making it more and more efficient, and we do not have the pretention to replace them in one paper.\\

It is most probable that any algorithm we propose would be a lot further from reality than the model in which the market maker is informed, because the market naturally tends to this optimal state. Therefore, we do not claim to know what the market impact really is (because it certainly changes as the actors change their strategies). We only claim to know what it tends to be, when actors are more and more clever (which is certainly the case since our findings pretty well account for reality).\\

An interesting study though is to see what changes if we assume that information is asymmetric. Indeed, as it introduces risk for the market makers, their trading would become a trade-off between returns and risk, and they would no longer trade according to zero-profit strategies. This could also happen if competition was not perfect, that is to say if there were not enough arbitragers on the market to incite people finding more and more clever strategies to make profits. Then the informed traders would no longer trade at the fair price described in \cite{farmer5}, but at higher prices, and they would probably have more transient impact on the price (which would make it more unstable).\\

Let us try to quantify this loss. If informed traders trade at rate $\mu$, a meta-order of length $l$ occurs on a time $t\simeq\frac{l}{\mu\overline{n}}$ where $\overline{n}$ is the average size of a trade (this allows to have $t$ in trade time instead of unit volume time). With naive estimators, if $N_{nt}$ is the average number of trades from noise traders during time $t$, the imbalance in order flow due to noise traders after time $t$ (which is also the uncertainty on $l$) is of order

\bea
\Delta l\approx\overline{n}\sqrt{N_{nt}}&\approx& \overline{n}\left(\frac{l_{nt}}{\overline{n}}\right)^{1/2}\nonumber\\
&\approx&\overline{n}\left(\frac{1-\mu}{\mu}\frac{l}{\overline{n}}\right)^{1/2}\nonumber\\
&\approx& \sqrt{\frac{1-\mu}{\mu}}(\overline{n}l)^{1/2}
\eea

Therefore, the uncertainty on the permanent market impact (i.e. the difference between the average trading price the price after reversion, which must be set by the market maker to counterbalance the uncertainty) is of order

\bea
\Delta p_{\infty}(l) &\approx& \frac{C}{\delta+1}\left(l+ \sqrt{\frac{1-\mu}{\mu}\overline{n}l}\right)^{\delta}-\frac{C}{\delta+1} l^{\delta} \nonumber\\
&\approx& \frac{C}{\delta+1} l^{\delta}\left[\left(1+\sqrt{\frac{(1-\mu)\overline{n}}{\mu l}}\right)^{\delta}-1 \right] \nonumber\\
&\approx& \frac{C \delta }{\delta+1}\sqrt{\frac{(1-\mu)\overline{n}}{\mu}} l^{\delta-1/2}\nonumber\\
&\approx&\delta\sigma_{T} \frac{\sqrt{1-\mu}}{\mu} \frac{l^{\delta-1/2}}{\EE\left(l^{2\delta-1}\right)}
\eea

\begin{remark}
For the common values of the correlation exponent $\gamma$, this is roughly constant (recall that $\delta=\gamma-1\approx1/2$).
\end{remark}

\begin{remark}
$\Delta p_{\infty}$ can be regarded as an inefficiency prime per trade which would be paid by the informed traders.
\end{remark}

For a meta order of length $l$, the total loss is of order

\bea
l\Delta p_{\infty} &\approx& \delta\sigma_{T} \frac{\sqrt{1-\mu}}{\mu}\frac{l^{\delta+1/2}}{\EE\left(l^{2\delta-1}\right)}\nonumber\\
&\approx&\delta\sigma_{T}\frac{\sqrt{1-\mu}}{\mu} l \quad \text{for $\delta\approx0.5$}
\eea

The constants being of order 1, we see that for any $l$ the additional cost is a few ticks per trade (which is non-negligible). This computation allows one to estimate the collective gain in improving arbitrage strategies (market makers have less risk, informed traders execute at better prices). Of course, we considered here in the worst case where the market maker does not make a difference at all between noise and informed traders, so the loss may be overestimated - but it gives the order of magnitude, which is the tick. One interpretation of these few ticks is that they are gained by setting a bigger spread.\\

One could argue that markets become meaningless if middle men trade a too large part of the volume. But, if they are clever enough (competitive enough), they only react to informed trades (and can estimate how much information a trade contains). In our ideal model, at least half of the trades are made by middle men, and nevertheless only informed trades change the prices. In any way, they have no interest in moving the price (apart from trying to set it as close as possible to the ideal price, which seems to contributes only to stability).\\

\section{Execution costs}\label{sec:OptimalExecution}

\subsection{Fast informed trading}\label{sec:fastIT}

The previous results and the ideas exposed above allow at least a qualitative interpretation (before going further and computing optimal strategies). There is what we could call an imperfection in the market which comes from asymmetry of information between the informed trader and the market maker: the decay to $p_{\infty}$. This may be exploited by the informed trader to have a better execution price.\\

As there is no such thing as a continuous trading (since there is a minimal unit of volume), we understand that an optimal strategy may be non-trivial. Indeed, take a meta-order of volume 100, a liquidation in one large order is exactly the same than a fast liquidation in 100 buckets. Thus there must be an optimal size for the buckets to minimize the impact.\\

Let us assume that the informed trader is still slow enough in comparison to noise traders and market makers, so that they have the time to react. Let us also assume here that he is the only informed trader while he executes his trade (so that the decay of market impact totally depends on his behavior, which is approximately true if he is fast and big enough). If the market impact function is scale invariant as we have assumed above, the price reaches $p_{\infty}$  at time $t+t_{max}=(1+\lambda) t_{max}$, where $\lambda$ depends only on the correlation exponent $\gamma$ (or more generally on the correlation function of trades). The gain for the informed trader if he resumes trading at that time is also the loss for the market maker $\Delta_{1}(p_{\infty})$, as defined by eq.\ref{eq:loss1}:

\bea
\Delta_{1}(p_{\infty}) = \int _{0}^{L_{eq}} I(L+l) dl - \int _{0}^{L_{eq}} (p_{\infty}+I(l))dl
\label{eq:gain}
\eea

where, as before, $I(l)$ is the transient market impact function and $L_{eq}$ is the volume of trades needed to reach the correct price again, after complete decay to $p_{\infty}$. We now know that $p_{\infty}=\frac{p_{max}}{\delta+1}$, $I(l)=Cl^{\delta}$ and also that $p_{\infty}+I(L_{eq})=I(L+L_{eq})$ by definition. Therefore, eq.\ref{eq:gain} becomes

\bea
\Delta_{1}(p_{\infty}) = \frac{C(L+L_{eq})^{\delta+1}}{\delta+1}-\frac{CL^{\delta+1}}{\delta+1} - L_{eq}p_{\infty}-\frac{CL_{eq}^{\delta+1}}{\delta+1}
\eea

which simplifies in

\bea
\Delta_{1}(p_{\infty}) = \frac{CL^{\delta+1}}{\delta+1}((1+\frac{L_{eq}}{L})^{\delta+1}-1 - \frac{L_{eq}}{L}-(\frac{L_{eq}}{L})^{\delta+1})
\label{eq:gain2}
\eea

where $L_{eq}$ is defined by

\bea
p_{\infty}=I(L+L_{eq})-I(L_{eq})
\eea

which reads, with the known expression of $I(l)$ and $p_{\infty}$,

\bea
\frac{1}{\delta+1}=(1+\frac{L_{eq}}{L})^{\delta}-(\frac{L_{eq}}{L})^{\delta}
\eea

Therefore $\frac{L_{eq}}{L}=f(\delta)$, and if we define $g(\delta)=(1+f(\delta))^{\delta+1}-1 - f(\delta)-f(\delta)^{\delta+1}$, the expression \ref{eq:gain2} of the gain for the informed trader due to his strategy becomes

\bea
\Delta_{1}(p_{\infty}) = g(\delta) \frac{C}{\delta+1} L^{\delta+1}
\label{eq:gainOneTrade}
\eea

With the basic strategy, the cost of market impact would be $\mathcal{C}=\frac{C}{\delta+1} (L+L_{eq})^{\delta+1}$. The relative gain is then

\bea
\frac{\Delta_{1}(p_{\infty})}{\mathcal{C}}=\frac{g(\delta)}{(1+f(\delta))^{\delta+1}} =\mathcal{G}(\delta)
\eea

\begin{figure}[h]
\begin{center}
\includegraphics[scale=0.45]{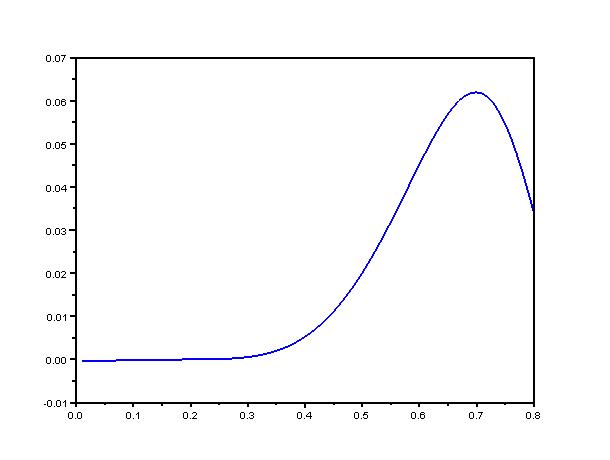}
\caption{The relative cost saving $\mathcal{G}(\delta)$ due to the decay of market impact.}
\label{fig:cprime}
\end{center}
\end{figure}

 \begin{remark}
 $\mathcal{G}(\delta)$ is of order 0.02 to 0.06 depending on $\delta$ (cf. Fig.\ref{fig:cprime}). This means that the gain can be a fixed fraction of the cost of market impact (which is not negligible).\\
\end{remark}

Let us finally consider the following strategy, represented on fig.\ref{fig:stratOpt} (which may not be optimal) to estimate the cost saving obtained by using an appropriate execution strategy: the informed trader begins with $l_{0}$ trades, waits for the decay to be complete, then sends $l_{1}=L_{eq}=f(\delta)l_{0}$ market orders (just so as to reach the normal price again), etc. If we note $l_{n}$ the size of the $(n+1)^{th}$ bucket, and $L_{n}=\sum_{i=0}^{n}l_{i}$ the total volume submitted so far, the size of the $n^{th}$ buckets size is then:

\begin{figure}[h]
\begin{center}
\includegraphics[scale=0.6]{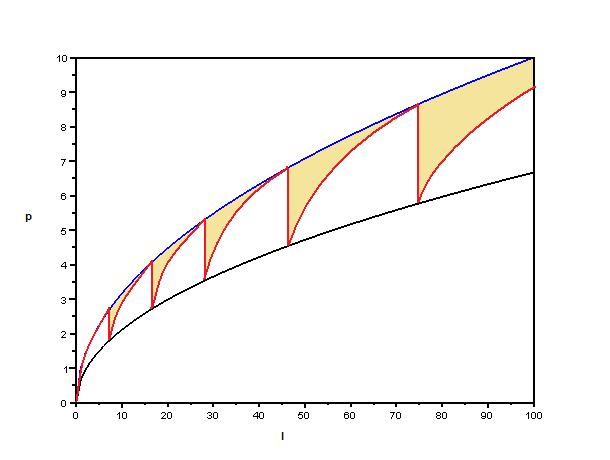}
\caption{For the strategy considered in the example, the price path is represented in red, in function of the volume submitted (note that the trader does not trade during the decay, which is thus immediate if the time unit is the volume submitted). The cost saving in comparison to a VWAP is represented in beige: each area is equal to $\Delta_{1}$ computed above, for the corresponding price.}
\label{fig:stratOpt}
\end{center}
\end{figure}

\bea
l_{n}=L_{eq}(L_{n-1})=f(\delta)L_{n-1}
\eea

so that the recursion relation for the volume submitted after $n$ steps is

\bea
L_{n}=(1+f(\delta))L_{n-1}
\eea

The volume executed after $N$ steps is therefore

\bea
L_{N} = L_{0}(1+f(\delta))^{N}
\eea

and the cost saving in comparison to the the cost of the naive strategy $\frac{C L^{\delta+1}}{\delta+1}$, using formula \ref{eq:gainOneTrade} for each step, is

\bea
\Delta_{strat, N} &=& \sum_{n=0}^{N-1} \Delta_{1}(p_{\infty}(L_{n})) \nonumber\\
&=& \frac{C}{\delta+1} L_{0}^{\delta+1}g(\delta) \sum_{n=0}^{N-1} (1+f(\delta))^{n(\delta+1)}\nonumber\\
&=&\frac{C}{\delta+1} L_{N}^{\delta+1}g(\delta) \sum_{n=1}^{N} (1+f(\delta))^{-n(\delta+1)}\nonumber\\
&=& \frac{C}{\delta+1} L_{N}^{\delta+1} \mathcal{G}(\delta)\frac{1-\left(1+f(\delta)\right)^{-N(\delta+1)}}{1-(1+f(\delta))^{-(\delta+1)}}
\eea

\begin{figure}[h]
\begin{center}
\includegraphics[scale=0.45]{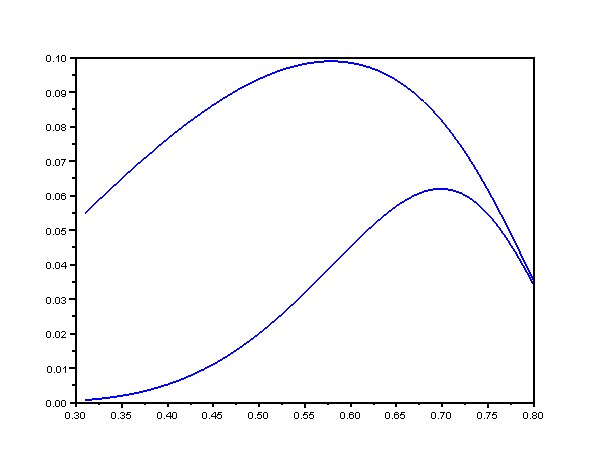}
\caption{Plot of $\mathcal{G}(\delta)$ and $\mathcal{G}_{max}(\delta)$. The cost savings with the strategy proposed is of order 10\% of the market impact cost. The actual savings are between the curves if $N$ is finite, and increase with $N$.}
\label{fig:gainStrat}
\end{center}
\end{figure}

\begin{remark}
If we note $\mathcal{G}_{max}(\delta)=\frac{\mathcal{G}(\delta)}{1-(1+f(\delta))^{-(\delta+1)}}$, then for any $N$ the relative cost saving belongs to $[\mathcal{G}(\delta),\mathcal{G}_{max}(\delta)]$, where $\mathcal{G}(\delta)$ is the cost saving when only one step is realized ($N=1$) and $\mathcal{G}_{max}(\delta)$ is the cost saving if the number of steps is infinite (which is purely theoretical since volumes cannot be smaller than the unit volume). $\mathcal{G}(\delta)$ and $\mathcal{G}_{max}(\delta)$ are represented on Fig.\ref{fig:gainStrat}.
\end{remark}

This means that an optimal liquidation using variable trading rate can decrease the cost of market impact up to 10\% (at least, since the strategy above may be suboptimal) for $\delta$ of order $0.5$.\\

Let us finally consider execution times. According to the strategy presented above, a meta-order in linear time, since $L_{N}$ market orders are submitted in time

\bea
T_{N}&=&\sum_{k=0}^{N}L_{N}^{k} \lambda^{N-k}\nonumber\\
&=& \lambda^{N}\sum_{k=0}^{N}\left(\frac{1+f(\delta)}{\lambda}\right)^{k}\nonumber\\
&=&\frac{\left(1+f(\delta)\right)^{N+1}-\lambda^{N+1}}{1+f(\delta)-\lambda}\nonumber\\
&\sim&\frac{1+f(\delta)}{1+f(\delta)-\lambda}L_{N} \quad \text{ if  $\lambda<1+f(\delta)$}
\eea

\begin{remark}
We found numerically that $\lambda\lesssim1$ in section \ref{sec:decay} for the common values of $\delta$, for the slowest possible decay process.
\end{remark}

\begin{remark}
A trivial strategy, consisting in executing one trade and wait for the decay before executing the next one, would have a much better gain since it would almost allow one to trade at price $\frac{p}{\delta+1}$ instead of $p$ for each trade and save 30 to 40\% of the market impact cost depending on $\delta$ (note that it is an upper bound to
the potential savings). But the completion time of this strategy would be quadratic in the volume:

\bea
T'\sim \frac{\lambda L^{2}}{2}
\eea

\end{remark}

\subsection{Slow informed trading: the cost of VWAPs}

Let us now consider the case where the informed trader does not represent a large part of the overall informed trading (which is the case for small actors or if the trading rate is small enough). In this case, the decay of market impact is not directly due to his behavior, but to the behavior of the other informed traders. We have shown in section \ref{sec:MIindiv} that in this case, the smaller the trading rate the smaller the impact. What about the cost of the strategy?\\

Let us introduce again the meta-order size $l_{A}$ of trader A, the number of periods during which the global trend is constant $N$, the volume executed by trader A during the $i^{th}$ period $\Delta l_{i}$, and the volume executed by trader B (the rest of the market) $l_{i}$. Let us assume that the trading rate is constant during each period. Due to the perfect competition, the average cost of the $\Delta l_{i}$ trades is equal to the permanent market impact $I(l_{i}+\Delta l_{i})=\frac{C}{\delta+1}(l_{i}+\Delta l_{i})^{\delta}$. Noting $\epsilon_{i}$ the sense of the global trend, if the initial price is set to zero,  then the price of the $\Delta l_{i}$ trades is

\bea
p_{i}=\frac{C}{\delta+1}\sum_{k=0}^{i}\epsilon_{k}(l_{k}+\epsilon_{k}\Delta l_{k})^{\delta}
\eea

since it is equal to the permanent impact of the previous trades plus the marginal cost of the $\Delta l_{i}$ trades. Therefore the total cost of the strategy is:

\bea
\mathcal{C}=\frac{C}{\delta+1}\sum_{i=0}^{N}\Delta l_{i}\sum_{k=0}^{i}\epsilon_{k}(l_{k}+\epsilon_{k}\Delta l_{k})^{\delta}
\eea

Finding an optimal execution strategy amounts to minimizing this quantity with respect to the $\Delta l_{i}$.

\begin{remark}
The quantities $l_{i}$ are not known beforehand, therefore the optimal strategy may be dynamical.
\end{remark}

Let us determine the cost of rapidity for constant trading rates $\widetilde{\mu}$ for trader A and $\mu$ for trader B. The expression of $\Delta l_{i}$ is given by \ref{eq:li}. Then

\bea
\mathcal{C}=\frac{C}{\delta+1}\frac{\widetilde{\mu}}{\mu}\sum_{i=0}^{N}l_{i}\sum_{k=0}^{i}\epsilon_{k}\left(1+\epsilon_{k}\frac{\widetilde{\mu}}{\mu}\right)^{\delta}l_{k}^{\delta}
\label{eq:costExact}
\eea

If $\widetilde{\mu}\ll\mu$,

\bea
\mathcal{C}=\frac{C}{\delta+1}\frac{\widetilde{\mu}}{\mu}\sum_{i=0}^{N}l_{i}\sum_{k=0}^{i}\epsilon_{k}\left(1+\epsilon_{k}\delta\frac{\widetilde{\mu}}{\mu}\right)l_{k}^{\delta}
\label{eq:costApprox}
\eea

If we make the approximation that $\EE(\epsilon_{k}l_{k}^{\delta})=0$, since each period is independent (by definition of our model), the expected cost is

\bea
\EE(\mathcal{C})=\frac{C\delta}{\delta+1}\left(\frac{\widetilde{\mu}}{\mu}\right)^{2}\EE\left(\sum_{i=0}^{N}l_{i}\sum_{k=0}^{i}l_{k}^{\delta} | \sum_{i=0}^{N}l_{i} = \frac{\mu}{\widetilde{\mu}}l_{A}\right)
\eea

\begin{figure}[h]
\begin{center}
\includegraphics[scale=0.45]{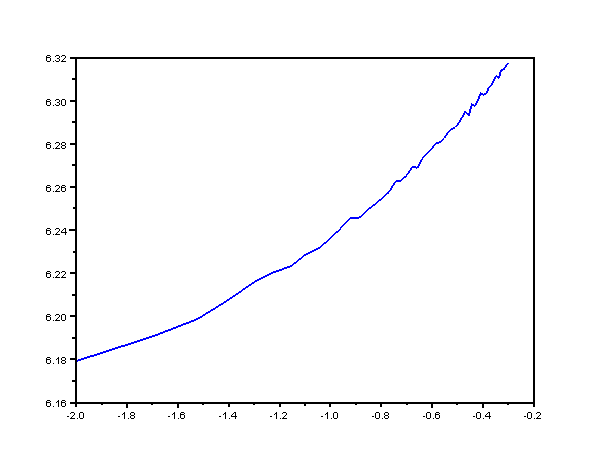}
\caption{The expected cost of the strategy $\EE(\mathcal{C})$ with constant rate $\widetilde{\mu}$, in function of $\frac{\widetilde{\mu}}{\mu}$ (log-log scale). The constant $\beta \in (0.1,0.2)$ for $\delta=0.5$.}
\label{fig:costStrat}
\end{center}
\end{figure}

\begin{remark}
In the particular case where there is only one period (either because $l_{A}$ is very small or $\widetilde{\mu}$ is large, or because $l_{0}$ happens to be large),

\bea
\EE(\mathcal{C})\sim \widetilde{\mu}^{1-\delta} l_{A}^{\delta+1}
\eea

The cost of trading is here roughly proportional to the square root of the participation rate. The opposite situation would be the case where there would be exactly $l_{A}$ periods of the same size (trader A would then trade a unit volume in each period). Then

\bea
\EE(\mathcal{C})\sim l_{A}^{2}
\eea

 and the cost would not depend on the participation rate. In the reality (with no approximation for eq.\ref{eq:costExact}), one would expect

\bea
\EE(\mathcal{C})\sim \widetilde{\mu}^{\beta(\widetilde{\mu})} \text{where $0<\beta(\widetilde{\mu})<1-\delta$ and is increasing.}
\eea

Numerical results are shown on fig.\ref{fig:costStrat}.

\end{remark}

\begin{remark}
The average cost of trading gets smaller when the participation rate $\widetilde{\mu}\rightarrow0$. However, the optimal execution strategy is a trade-off between the expectation of the cost and the variance (which grows with $\widetilde{\mu}$). It is then immediate to find general solutions for given $\beta$.
\end{remark}

The variance of the cost can be computed from \ref{eq:costApprox}, by assuming that $\epsilon_{k}$ and $l_{k}$ are independent:

\bea
\EE(\mathcal{V})=\left(\frac{C}{\delta+1}\right)^{2}\left(\frac{\widetilde{\mu}}{\mu}\right)^{2}\EE\left( \sum_{i=0}^{N}l_{i}^{2}\left[ \left( \sum_{k=0}^{i}l_{k}^{\delta}\right)^{2}+\frac{\delta^{2} \widetilde{\mu}^{2}}{\mu^{2}} \sum_{k=0}^{i}l_{k}^{2\delta}\right] \right)-\EE(\mathcal{C})^{2}
\eea

Again, the extreme cases are $l_{0}=l_{A}$ where $\mathcal{V}\sim \frac{l_{A}^{3}}{\widetilde{\mu}}$ and $l_{i}=1$ where $\mathcal{V}\sim\frac{l_{A}^{2+2\delta}}{\widetilde{\mu}^{2\delta}}$. For $\delta\simeq1/2$, the variance always behaves like

\bea
\EE(\mathcal{V})\sim C^{2}\frac{l_{A}^{3}}{\widetilde{\mu}}
\eea

\begin{remark}
The variance  for a pure brownian motion would be

\bea
\EE(\mathcal{V}_{b})&=&\EE\left[\left(\int_{0}^{\frac{l_{A}}{\widetilde{\mu}}}\widetilde{\mu}t \sigma dW_{t}\right)^{2}\right]
=\int_{0}^{\frac{l_{A}}{\widetilde{\mu}}}\left(\widetilde{\mu}t\right)^{2} \sigma^{2} dt =\frac{\sigma^{2}l_{A}^{3}}{3\widetilde{\mu}}
\eea

and would have the same qualitative dependance in the parameters (recall that $C\sim\sigma$).

\end{remark}

In the case of a constant-rate execution, the optimal static strategy is therefore the result of the classical mean-variance optimization:

\bea
\widetilde{\mu}&=&argmin\left( \EE(-\mathcal{C})-\lambda \EE(\mathcal{V}) \right)
\eea

where $\lambda$ represents the risk-aversion of the informed trader.

\begin{remark}
The previous reasoning can be made for variable $\widetilde{\mu}$, in order to find optimal dynamical strategies.
\end{remark}

\section{Summary and comments}

The framework that we set in our model has the following nice properties:

\begin{itemize}
\item Prices are discrete, so any real quantity exists in our framework.

\item Consistent with economic theory, information is incorporated into prices thanks to a non-zero permanent impact. The informational content of trades varies though, as more predictable orders have less impact.

\item Prices are martingales even though the order flow is autocorrelated and private information is incorporated. Concerning public information, one could add a jump process which would not affect our results.

\item In spite of a concave impact function, prices are not manipulable. This comes from the fact that the fundamental impact is the impact of the aggregated order flow and not the renormalized impact: it is the way the agents fit into the market that defines their impact. Small meta-orders being essentially executed during one only trend, they have a concave impact, but large meta-orders are executed while the background trends change - the memory of the market in erased in some sort on that time scales, so that impact becomes additive and thus linear. The concavity on short time scales can be viewed as a premium paid for fast execution.

\item It provides a natural interpretation for the latent order book - and one could also find the shape of the displayed order-book.

\item The competition between market makers, arbitragers, and their "intelligence", are present in what we called the market maker: his response to the order flow - coming from classical investors or news traders - consists in his limit orders placing strategy. Any kind of statistical arbitrage should be seen as intentions to get closer to the ideal price path. The benefits of such strategies, in a market where competition is not perfect, can be computed.

\item It explains many empirical results about the renormalized impact, the aggregated market impact, the cost of VWAPs.

\item It explains why the observed impact, transient or permanent, varies among traders, and shows how the quality of a strategy can be deduced from the observed permanent impact.

\end{itemize}

A few things have not been studied in this paper - whereas the framework would allow it. The most important is certainly the spread, which has no reason to emerge here since the adverse selection of private information is what defines the strategy of the market-maker. To study the market impact, we did not need to consider public information. A way to integrate it would be to add a jump process to our price process, to which we could associate an adverse selection. The spread would emerge naturally as a compensation for adverse selection.\\

This leads to the last reflection of this paper. Until here, we have considered the trends - or what we call the informed trader in the first sections - uniquely from a statistical point of view, which was enough for our study because no other information is available to the rest of the market - what we called the market maker. But now if we consider the filtration which contains the private information of the informed trader, then - now assuming perfect competition between the informed traders, news arbitragers, etc - he must behave so that his permanent impact is the expectation of the price conditioned on his private information. In other words, the (informed) order flow must be such that the additional jump process corresponding to public information is a martingale. It is now possible to understand the noise trader - apparently an absurd agent until here - who comes from the heterogeneity of private information between the informed traders. So it appears that people who trade with the less accurate information tend to lose money... \\

Stated this way, the notion of price manipulation appears in a brand new light. If some trader pushes the price up with no underlying justification, his trades will then be anti-correlated to the jump process - which will no longer be a martingale. And indeed, even without any notion of fundamental value of the price, one senses that if a buy trend was not followed by any good public information, the price would eventually come back to its original level. Price manipulation, in this context, could only be achieved by momentum trading which would only shift the price while it occurs - and this could only appear for huge volumes because of the argument that immediately follows. In any case, someone who would try to manipulate the price would be arbitraged - and lose money.\\

The last question will then be: what is the difference between the \{public+private information\} process, and the \{public information\} process. Indeed, since one trader's information becomes a piece of information for all the other agents, one could sense that the former process may be more unstable than the latter. This kind of collective behavior has been studied in detail (\cite{collective} for example).

\section{Conclusion}

We have shown that the zero-profit condition for a market-maker resulting in perfect competition - and the martingale condition - leads to an optimal market-making strategy in a market in which the order flow is correlated. In a particular case, we have shown that it imposes a particular shape for the \textit{Bare Market Impact} function, and gives a relation between its shape while the trend is occurring and the price after the completion (which is non-zero), and the correlations in order flow present in the market. We have made the computation for the empirical shape of the order flow correlations found in \cite{eisler}, which leads asymptotically to a power law market impact. We have found general expressions for the \textit{Renormalized Market Impact} function and provided a way to compute it in any particular situation. In particular, for power-law correlations in the market, it is found to be power-law for small volumes and to become progressively linear for large volumes, thus forbidding price manipulation. We have seen that the different strategies for submitting a meta-order have different costs, due partly to the decay after the trends and partly to the trading rate. In the case of a VWAP, we have expressed this cost explicitly, and shown that it increases as a power-law of the trading rate, with exponent $\sim 0.1-0.2$. In introducing the notion of public information, we have finally shown why the competition between informed traders definitively rules out any price manipulation.

To conclude, let us ask the reverse question: why does reality looks like what we find here. Indeed, in our whole study we have assumed that market makers were perfectly informed and competitive, and therefore the shape of market impact that we have found is the average shape that we would find in ideal markets: the similarity to real data actually suggests that markets are not far from being ideal from this point of view.\\

\section*{Acknowledgements}

I would like to thank Jim Gatheral for his availability and his support. None of this could have emerged without all he taught me.

I would also like to give special thanks to Antoine Tilloy and Pierre Baqué for their very useful comments and always enlightening discussions.

\bibliographystyle{plain}
\bibliography{bibli2}

\end{document}